\documentclass{PoS}
\usepackage{aas, longtable, caption}

\title{Recent outburst activity of the symbiotic binary \\AG Draconis}

\ShortTitle{Recent outburst activity of the symbiotic binary AG Draconis}

\author{\speaker{Jaroslav Merc}\\
        Faculty of Science, P. J. \v{S}af{\'a}rik University, Park Angelinum 9, 040 01 Ko\v{s}ice, Slovak Republic\\
        E-mail: \email{jaroslav.merc@student.upjs.sk}}

\author{Rudolf G{\'a}lis\\
        Faculty of Science, P. J. \v{S}af{\'a}rik University, Park Angelinum 9, 040 01 Ko\v{s}ice, Slovak Republic\\
        E-mail: \email{rudolf.galis@upjs.sk}}

\author{Laurits Leedj{\"a}rv\\
        Tartu Observatory, University of Tartu, Observatooriumi 1, T\~{o}ravere, 61602 Tartumaa, Estonia\\
        E-mail: \email{laurits.leedjarv@to.ee}}

\abstract{The symbiotic binary AG Dra regularly undergoes quiescent and active stages which consist of several outbursts repeating with about 360\,d interval. The recent outburst activity of AG~Dra started by the minor outburst in the late spring of 2015 and was definitely confirmed by the outbursts in April 2016 and May 2017. In the presented work, the photometric and spectroscopic behaviour of the recent outburst activity of AG~Dra is presented in detail. Moreover, the temperature of the white dwarf in AG~Dra is studied based on the behaviour of the prominent emission lines. We show that a disentanglement of particular effects in the observed changes of the emission lines is crucial to investigate the intrinsic white dwarf temperature variations related to outburst activity of this strongly interacting binary. We also report the effects of the low excitation lines orbital variations and of the \mbox{H$_{\beta}$} absorption component on their equivalent widths as well as consequences of the approximations used in our previous works.}

\FullConference{The Golden Age of Cataclysmic Variables and Related Objects IV\\
                 11-16 September, 2017\\
                 Palermo, Italy}

\begin{document}

\section{Introduction}

AG~Dra is a long-known symbiotic variable star with the orbital period around 550 days (Meinunger, 1979; G{\'a}lis et al., 1999; Hric et al., 2014). This binary system consists of a metal-poor red giant of spectral type K3\,III (Shenavrin et al., 2011) with 
mass of 1.5\,M$_{\odot}$ (Kenyon \& Fernandez-Castro, 1987). The giant is under-filling its Roche lobe (Sion et al., 2012) and the accretion most likely takes place by the stellar wind from the cool giant.

The hot component of AG~Dra is a white dwarf (WD) with a temperature of (1\,-\,1.5)\,$\times\,10^{5}\,$K (Miko{\l}ajewska et al., 1995; Shore et al., 2010) and luminosity of (1\,-\,5)\,$\times\,10 ^{3} $\,L$_{\odot}$. The WD's mass is approximately estimated as (0.4\,-\,0.6)\,M$_{\odot}$ (Miko{\l}ajewska et al., 1995). Due to the giant's wind, the binary is surrounded by an extended circumbinary nebula, partially ionised by the WD. It remains unclear whether an accretion disk surrounding the hot component exists or not in this symbiotic system.

AG~Dra is a non-eclipsing binary whose inclination is proposed to be \mbox{40$^{\circ}$ - 70$^{\circ}$}. The distance of AG~Dra is quite uncertain: the Hipparcos satellite has set a lower limit of 1 kpc; Miko{\l}ajewska et al. (1995) and Sion et al. (2012) proposed the distance of 2.5\,kpc and 1.5\,kpc, respectively. Observations of the Gaia satellite will probably be sufficiently clear\footnote{Note added in revision: In April 2018, the parallax of $(0.210 \pm 0.027)$\,mas was published for AG~Dra in the Gaia DR2 (Gaia Collaboration et al., 2018). The reliable distance though cannot be obtained by inverting the parallax (Luri et al., 2018). Using inference procedure taking to account for the nonlinearity of the transformation and the asymmetry of the resulting probability distribution, Bailer-Jones et al. (2018) derived for AG~Dra the point distance of 3.89\,kpc with uncertainty of (3.53\,-\,4.32)\,kpc representing $\approx$\,68\% confidence interval.}. Its low metallicity, high spatial velocity ($v_{\rm r}$ = -148 km\,s$^{-1}$) and high Galactic latitude ($b = 41^{\circ}$) suggest that AG~Dra belongs to the old halo population (Schmid \& Nussbaumer, 1993).

The system manifests a characteristic symbiotic activity by showing active and quiescent stages alternatively (Kenyon, 1986). Individual outbursts during active stages of AG~Dra are repeating at approximately a one-year interval. The amplitude of the outbursts is from 1\,-\,1.4\,mag in the $V$ filter to 3.6\,mag in the $U$ filter (Leedj{\"a}rv et al., 2016). The active stages occur in intervals of 9\,-\,15 years (in 1936, 1951, 1966, 1980, 1994, 2006 and 2015; see figure 1 in Merc et al., 2017).

Gonz{\'a}lez-Riestra et al. (1999) showed that two types of outbursts are presented in AG~Dra which differ from each other in the hot component temperature. Major outbursts at the beginning of active stages (e.g. 1981-83, 1994-96 and 2006-08) are usually \textit{cool}, during which the pseudo-atmosphere of the WD cools down while its radius increases by the factor of 2 to 6 (Leedj{\"a}rv \& Burmeister, 2012). During smaller scale \textit{hot} outbursts, the temperature increases or it remains unchanged. In our previous work (Leedj{\"a}rv et al., 2016), we showed that the \textit{cool} and \textit{hot} outbursts of AG~Dra could be clearly distinguished by the behaviour of emission lines in optical spectra of this symbiotic system. 

In the paper, the photometric and spectroscopic behaviour of the recent activity of AG~Dra is analysed with aim to infer the type of individual outbursts. The ongoing activity stage is compared with ones previously recorded during almost 130 years of photometric observations of this symbiotic binary. Moreover, the temperature of the WD in AG~Dra is studied based on the behaviour of the prominent emission lines with focus on particular effects that can influence its determination.

\section{Observations}

Spectroscopic observations of AG~Dra acquired from \textit{Astronomical Ring for Access to Spectroscopy database} (ARAS)\footnote{http://www.astrosurf.com/aras/Aras\_DataBase/Symbiotics.htm} were analysed in this study. We used 101 spectra obtained between JD~2\,457\,110 (March 28, 2015) and JD~2\, 458\,044 (October 17, 2017). Even the spectra were acquired with small telescopes (25\,-\,35\,cm, $R = 1\,800\,-\,11\,000$), they provided us valuable information about the recent activity of AG~Dra. In addition, we analysed the intermediate-dispersion spectroscopy of AG~Dra covering previous stages of its activity, carried out at the Tartu Observatory in Estonia. Altogether, 515 spectra obtained during almost 14 yr (from JD~2\,450\,703 to JD~2\,455\,651) on the 1.5-m telescope ($R = 6\,000, 7\,000$ and 20\,000) were studied in our paper Leedj{\"a}rv et al. (2016). In this study, we reanalysed these data to investigate the temperature behaviour of the hot component of AG~Dra and to compare it with the recent activity of this symbiotic binary.

Our analysis was focused on the three prominent emission lines: the neutral helium \mbox{He\,{\sc i}} line at $\lambda$\,4471\,\AA, the ionised helium \mbox{He\,{\sc ii}} line at $\lambda$\,4686\,\AA\, and the hydrogen Balmer line  \mbox{H$_{\beta}$} at $\lambda$\,4861\,\AA. We measured equivalent widths (EWs) of these lines using the software PlotSpectra\footnote{https://goo.gl/caybSH} and computed the fluxes in lines using the photometric observations of AG~Dra obtained from \textit{AAVSO International Database} (Kafka, 2017) and Vra\v{s}\v{t}\'{a}k (2017).

\section{Recent outburst activity of AG~Dra}

After seven years of flat quiescence following the 2006-08 major outbursts, in the late spring of 2015, the symbiotic system AG Dra started to become brighter again toward what appeared to be a new minor outburst. The outburst activity of AG~Dra was definitely confirmed by the following two outbursts in April 2016 and May 2017\footnote{Note added in revision: In April 2018, the fourth minor outburst during the recent active stage of AG~Dra was observed (G{\'a}lis et al., 2018).}. 

\subsection{Photometric behaviour}

The first, less prominent outburst (G0) was observed in May 2015. The maximum brightness was achieved around JD~2\,457\,166 (10.7 and 9.6\,mag in the $B$ and $V$ filters, respectively). It turned out that it was minor outburst of AG~Dra, a precursor of its activity as it was observed in some of the previous active stages. 

During the second, more prominent outburst (G1), the brightness around  JD~2\,457\,517 (May 8, 2016) reached the maximum of 9.9 and 9.1\,mag in the $B$ and $V$ filters, respectively. As in the case of the previous outburst (G0), its amplitude ranks this outburst to the minor outbursts of AG~Dra. Such photometric behaviour of the active stage is very unusual. More often, the pre-outbursts of AG~Dra are followed by major outbursts, during which the brightness can reach around 8.8 and 8.4\,mag in the $B$ and $V$ filters, respectively.

According to our statistical analysis of photometric observations, we determined that the time interval between outbursts of AG~Dra vary from 300 to 400\,d (without an apparent long-term trend), with a median around 360\,d. We were able to expect the next outburst in the interval from May 3, 2017 (JD~2\,457\,877) to June 12, 2017 (JD~2\,457\,917) from this result. In April 2017, we have initialised the observational campaigns to study the photometric and spectroscopic behaviour of the recent active stage of AG~Dra\footnote{https://www.aavso.org/aavso-alert-notice-572}$^,$\footnote{ http://www.astrosurf.com/aras/Aras\_DataBase/Symbiotics/Campaigns/2017\_AGDra.pdf}. The system manifested the third outburst of the ongoing activity stage on May 17, 2017, 373 days after previous one. The maximum brightness of AG~Dra during this outburst was similar to the case of G0 and it was reached around JD~2\,457\,890. 

The photometric behaviour suggests that all three recent outbursts of AG~Dra belong to the \textit{hot} type. Such classification is also supported by the increase of the EWs of studied emission lines detected during all these events.

\subsection{Spectroscopic behaviour}

Spectroscopic behaviour of the recent outburst activity of AG~Dra was analysed using selected prominent emission lines detected in optical spectra of this interacting binary. To study basic properties of a symbiotic nebula, the emission lines of \mbox{He\,{\sc ii}} and \mbox{H$_{ \beta }$} are particularly interesting (Skopal et al., 2017). We used the ratio of their EWs to investigate the evolution of AG~Dra during outburst activity stages of this symbiotic system.

The emission lines have an origin in different parts of a symbiotic nebula, and these regions are different in their shape and size. Changes of their visibility due to binary revolutions cause variations of the emission line EWs with the orbital phase. Moreover, these EWs are variable during outbursts reflecting changes of properties of the symbiotic system components, e.g. the temperature of the WD as a source of ionising radiation. To study these intrinsic changes, it is necessary to carefully disentangle particular contributions to observed variations. 

In the following sections, we discuss a dependence on the orbital motion and other effects that can influence values of the EWs of the studied emission lines.

\subsubsection{Orbital dependence of the \mbox{He\,{\sc ii}} and H$_{\beta}$ ratio}

Photometric variations of AG~Dra during quiescence stages are the most pronounced in the $U$ filter with the amplitude around 0.5 mag and they decrease towards longer wavelengths. The contribution of the WD radiation should be not more than about 10\% in the $U$ filter (0.1 mag) during quiescence according to the temperatures and radii indicated by the results of Miko{\l}ajewska et al. (1995) and Greiner et al. (1997) based on analysis of ultraviolet and X-ray observations, respectively. It seems from these results and from the fact that quiescence variations in the $U$ filter are periodical with the orbital period that these variations can be explained by varying visibility of an ionised gaseous region, which is, in fact, the partially optically thick wind from the giant ionised by the WD in AG~Dra (Friedjung et al., 1998; Skopal, 2008). 
\begin{figure}[t]
	\vspace{-1.5cm}
	\includegraphics[width=\linewidth]{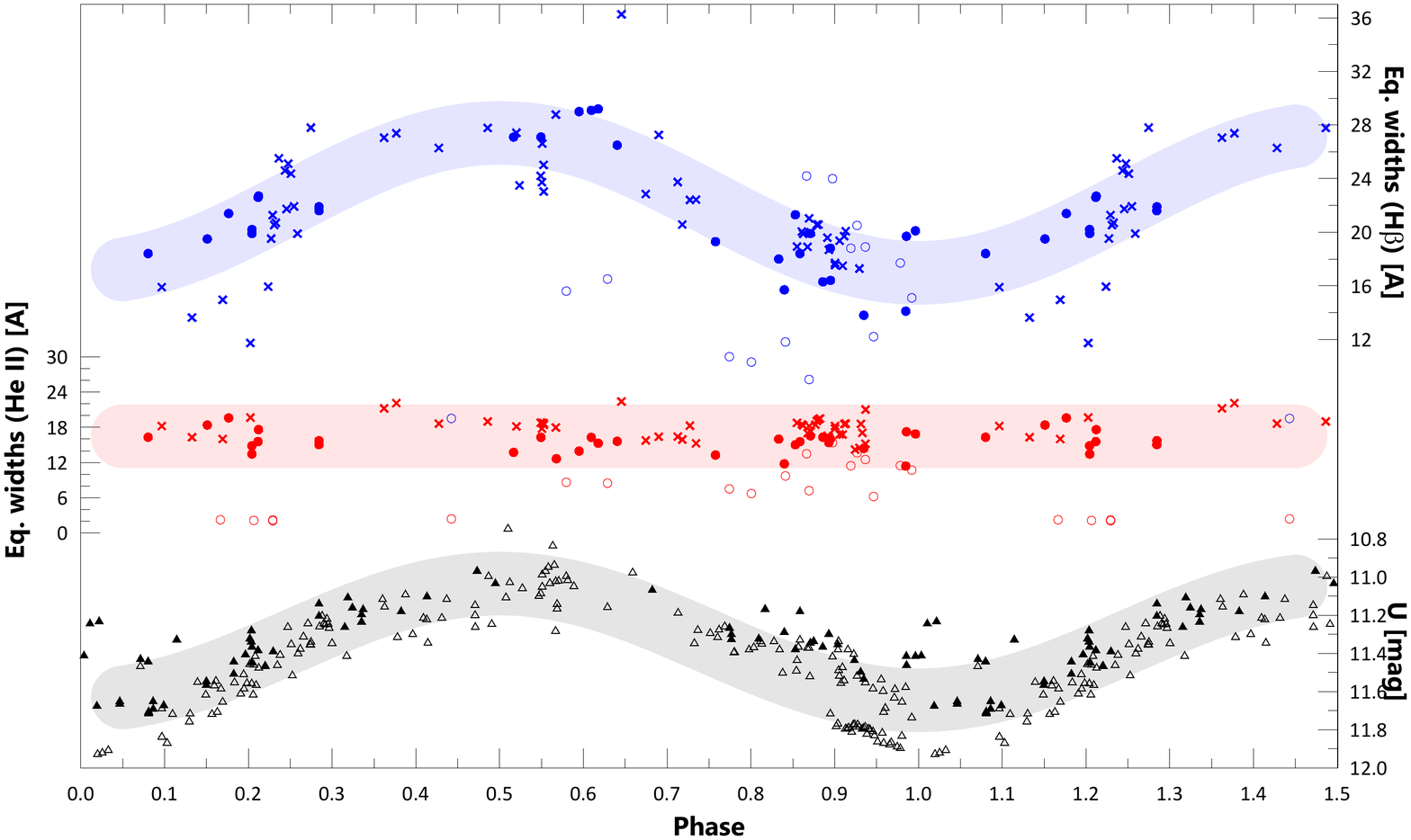}
	\caption{The orbital phase diagrams of the emission line EWs and the $U$ brightness of AG~Dra during the quiescence stages. Full triangles correspond to the possible quiescent episode between the active stages E and F (JD~2\,451\,200 - 2\,452\,100). Empty triangles represent observations obtained during the quiescent stage Q6 (JD~2\,454\,550 - 2\,457\,150). Circles and crosses represent data from Tartu Observatory and ARAS, respectively.}
	\label{fig:orbital_vari}
\end{figure}

According to our previous investigation (Hric et al., 2014; Leedj{\"a}rv et al., 2016), the similar orbital variations of the neutral hydrogen and neutral helium (at $\lambda$\,6678\,\AA) emission line EWs were also detected. These low excitation lines most likely arise in an extended gaseous volume which also emits continuum radiation in the near-UV and optical spectral region. At the same time, EWs of the high excitation \mbox{He\,{\sc ii}} $\lambda$\,4686\,\AA\, emission line practically do not vary with the orbital motion. This line should have its origin close to the hot component in AG~Dra.

The orbital phase diagrams of the He\,{\sc ii} $ \lambda $\,4686\,\AA\, and H$ _{\beta}$ EWs together with the $U$ brightness of AG~Dra during quiescence stages is depicted in figure \ref{fig:orbital_vari}. It is obvious that the \mbox{He\,{\sc ii}}/H$ _{\beta}$ ratio also varies with the orbital phase and these variations are in an anti-correlation with the photometric changes of AG~Dra during quiescence stages. Adopting the formula given by Iijima (1981), we used the ratio of the \mbox{He\,{\sc ii}} $ \lambda $\,4686\,\AA\, and H$_{\beta}$ emission line EWs as a proxy to the temperature of the hot component in AG~Dra (Leedj{\"a}rv et al., 2004, 2016; Merc et al., 2017). For this reason, the WD's temperature thus determined is also apparently changing with the orbital phase.

To investigate the intrinsic variation of the WD's temperature (e.g. due to the outburst activity of AG~Dra), it is necessary to disentangle particular effects in the observed variations of the \mbox{He\,{\sc ii}}/H$ _{\beta} $ ratio. We fitted the variation of the H$ _{\beta}$ and \mbox{He\,{\sc i}} $\lambda$\,4471\,\AA\, emission line EWs during quiescence by sinusoidal functions (with the fixed period given by the result of our analysis of radial velocities based on absorption line measurements). The EW increments as functions of the orbital phase for the studied emission lines are given by
\begin{eqnarray}
    \Delta EW_{H_{\beta}} &=& -5.228\,\cos(2\pi\,\varphi), \\
    \Delta EW_{4471} &=& -0.170\,\cos(2\pi\,\varphi).
\end{eqnarray}
The residuals are computed by subtraction of the obtained fits from corresponding EWs values. The average values of residuals are 1.30 and 22.19\,\AA\, for \mbox{He\,{\sc i}} and H$ _{\beta} $, respectively.
\begin{figure}[t]
	\vspace{-1.5cm}
	\includegraphics[width=\linewidth]{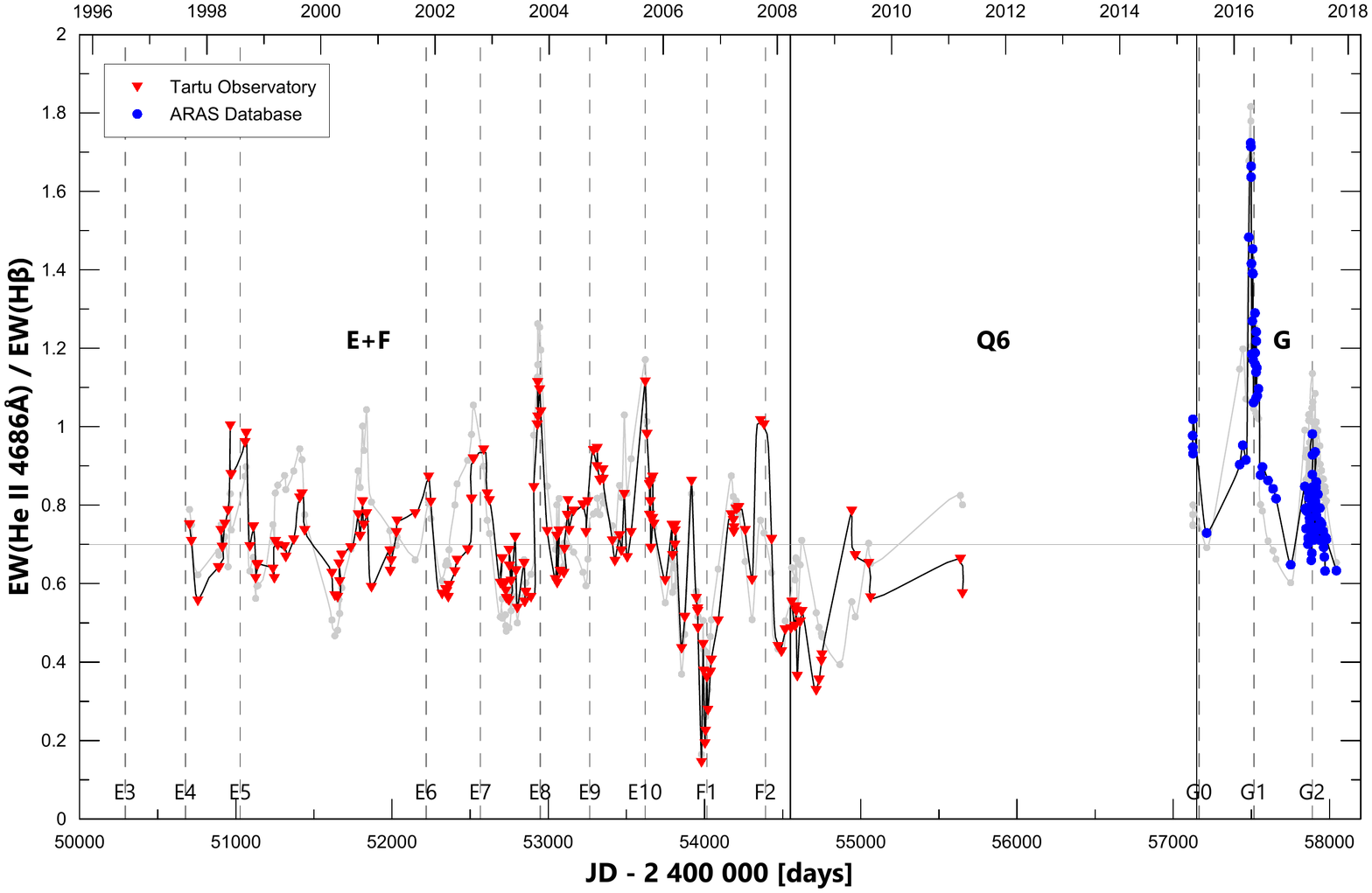}
	\vspace{-1.0cm}
	\caption{The ratio of EWs of two prominent emission lines H$_{\beta}$ and \mbox{He\,{\sc ii}} $ \lambda $\,4686\,\AA\, after subtraction of the orbital variations of H$_{\beta}$, in time. The data from Tartu Observatory and ARAS database are depicted by red triangles and blue circles, respectively. Original values of the ratio are depicted by the grey symbols. The vertical solid lines mark borders between the quiescence (Q6) and active (E+F, G) stages of AG~Dra. The vertical dashed lines highlight the individual outburst of this symbiotic binary. The long-term average of the \mbox{He\,{\sc ii}}/H$_{\beta}$ ratio over the period 1997-2011 with value of 0.7 is represented by the horizontal solid line.}
	\label{fig:temperature}
\end{figure}

In the next step, the residual EWs of H$ _{\beta}$ emission line are used to compute the rectified \mbox{He\,{\sc ii}}/H$ _{\beta} $ ratio values. These and original values of the ratio 
are depicted in figure \ref{fig:temperature}. Direct comparison of the rectified and original \mbox{He\,{\sc ii}}/H$ _{\beta} $ ratio curves showed a significant change of global behaviour of the ratio variations as well as dramatic changes of the ratio values for the particular outbursts of AG~Dra. It is worth to note that the amplitude of the \mbox{He\,{\sc ii}}/H$ _{\beta} $ ratio variations decreased for most of the outbursts except the cool one (F1), during which the ratio dropped to the value of 0.10 corresponding to the WD's temperature of 96\,000\,K. 

Other findings related to the rectification of the \mbox{He\,{\sc ii}}/H$_{\beta}$ values for the particular outbursts can be summarised as follows. (i) The possible cooling of the WD during and after E4 (based only on one observational point, which would be difficult to verify). (ii) The increase of the hot component temperature during E5, E6, E9 and G0. (iii) The outburst F2 (a part of double-peaked cool outburst F1-F2) is actually the \textit{hot} outburst. (iv) The dramatic drop of the WD's temperature during and after the recent outburst G2.

In the next section, the influence of a absorption component of the H$_{\beta}$ emission line on the estimation of its EWs and consequently on the \mbox{He\,{\sc ii}}/H$_{\beta}$ ratio values is discussed.  

\subsubsection{Absorption component of the H$_{\beta}$ emission line}

\begin{figure}[t]
	\vspace{-1.5cm}
	\includegraphics[width=\linewidth]{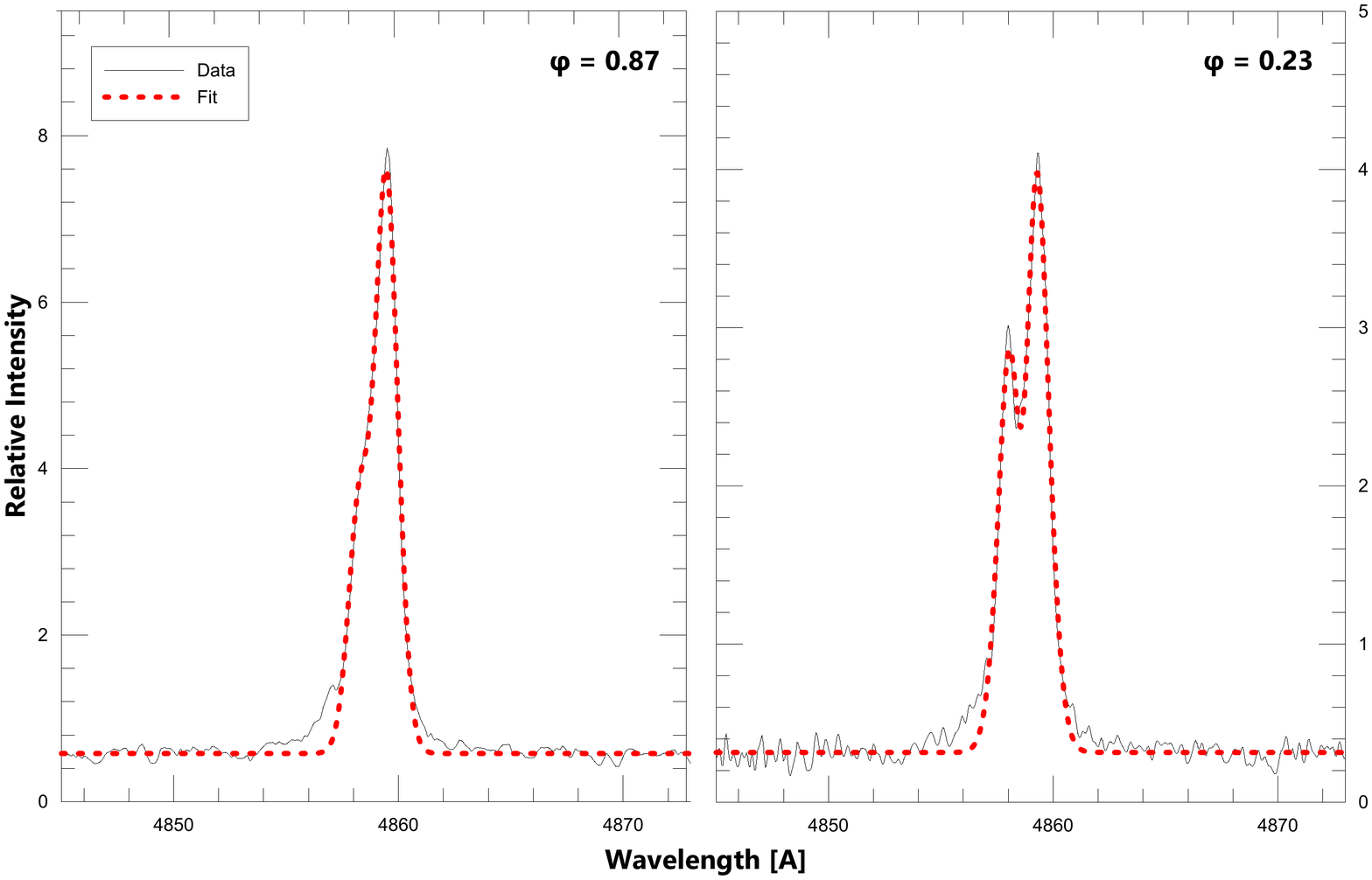}
	\caption{The profiles of the H$ _{\beta} $ emission line measured in different orbital phases (0.87 and 0.23 for left and right panel, respectively). Black lines represent original data; fit by two-term Gaussian is denoted by red dots. Phase zero corresponds to the inferior conjunction of the giant in AG~Dra.}
	\label{fig:hbeta}
\end{figure}

By detailed analysis of the emission lines, one can found that \mbox{He\,{\sc ii}} line profiles are more or less symmetrical, while profiles of the H$_{\beta}$ line are heavily affected by a absorption component in particular orbital phases of AG~Dra (figure \ref{fig:hbeta}). The presence of this component effectively reduces EWs of the H$_{\beta}$ line, resulting in an increase of the \mbox{He\,{\sc ii}}/H$_{\beta}$ ratio. This effect is particularly evident during and after the inferior conjunction of the giant and probably is related to the giant wind.

To estimate the total emission of the H$_{\beta}$ line in the case of AG~Peg, Skopal et al. (2017) removed the absorption component by fitting the line profiles using Gaussian curves. We adopted similar approach and fitted the emission line by two-term Gaussian to take into account both emission and absorption components. Analysis of 71 spectra with the resolution $R \approx$ 11\,000 revealed that the EWs of H$_{\beta}$ emission line could be reduced by 20\,-\,25\% in the case of AG~Dra. It causes the WD's temperature increase of about \mbox{10\%} if we used simplified method of Iijima (1981) described in the section \ref{temperature}. The analysis further indicated that absorption components are always presented in spectra of AG~Dra and cause the constant decrease of the H$_{\beta}$ EWs and therefore its effect does not change the course of observed variations.

Note that these results are influenced by the activity of AG Dra at a given time, and the significant emission during outburst can almost completely overlap the absorption component. For a detailed study of the dependence of the absorption component parameters on the orbital phase, we would need observational spectroscopic material to be obtained only during the quiescence of this symbiotic binary.

\subsection{Temperature of the WD}\label{temperature}

Determination of temperatures of central stars in planetary nebulae and of WDs in symbiotic systems is crucial in order to understand the essence of these systems. Zanstra (1931) proposed a method for obtaining the temperature of central stars in planetary nebulae using the intensities of Balmer lines and monochromatic magnitudes of the central star at the same frequencies. This method was later extended by Harman \& Seaton (1966) to account also the \mbox{He\,{\sc i}} and \mbox{He\,{\sc ii}} emission lines. Properties of hydrogen and helium emission lines alone can be used to determine the ionising source temperature as well. The first such method was proposed by Ambartsumyan (1932) and by Stoy (1933) and was modified by several authors (e.g. Kaler, 1976, 1978; Iijima, 1981; Kaler \& Jacoby, 1989). Using the Zanstra \mbox{H\,{\sc i}} and Zanstra \mbox{He\,{\sc ii}} temperatures, Kaler \& Jacoby (1989) derived a polynomial approximation for so-called "crossover temperature"
\begin{equation}
    \log (T_{\rm cross}) = 4.90500 + 1.11162\,\frac{F_{4686}}{F_{\rm H_{\beta}}}-1.10692\,\left(\frac{F_{4686}}{F_{\rm H_{\beta}}}\right)^2+0.62057\,\left(\frac{F_{4686}}{F_{\rm H_{\beta}}}\right)^3,
\end{equation}
which is valid for 0.08 < $F_{4686}$/$F_{\rm H_{\beta}}$ < 1.00. In our previous works (Leedj{\"a}rv et al., 2004, 2016; Merc et al., 2017), we used the method introduced by Iijima (1981), in which the temperature of the source of ionisation is calculated as
\begin{equation}\label{eq:origin}
T_{\rm hot}\,({\rm in\,10^4\,K}) = 19.38 \sqrt{2.22\,F_{4686} \over 4.16\,F_{\rm H_{\beta}}+9.94\,F_{4471} } + 5.13.
\end{equation}
The intermediate-dispersion spectra obtained at Tartu Observatory in the blue spectral region record only prominent emission spectral lines of \mbox{He\,{\sc ii}} and H$_{\beta}$ and do not include the \mbox{He\,{\sc i}} $\lambda$\,4471\,\AA\, line. From this reason, we were forced to introduce some approximations. In the next section, we discuss consequences of neglecting of the latter spectral line in our previous analysis. 

The lower limit of the ionising source temperature can by also obtained using the formula of Muerset \& Nussbaumer (1994). They found the simple empirical relationship between the temperature of the ionising source and the ionisation energy of atoms in particular ionisation state
\DeclareRobustCommand{\rchi}{{\mathpalette\irchi\relax}}
\newcommand{\irchi}[2]{\raisebox{\depth}{$#1\chi$}}
\begin{equation}
    T_{\rm hot}\,({\rm in\,10^4\,K}) = \frac{1}{10}\, \rchi_i,
\end{equation}
where $\rchi_i$ is the energy required to ionise atoms to the $i$-th ionisation state. According this formula, the presence of ions with the highest ionisation energy determines the lower limit for the ionisation source temperature, e.g. the minimal temperature for creation of \mbox{O\,{\sc vi}} lines is around 114\,000\,K. A~lack of five-time ionised atoms of oxygen when the temperature of the ionisation source was lower than this value resulted to the (almost) vanishing of the Raman scattered \mbox{O\,{\sc vi}} $\lambda$\,6825\,\AA\, emission line as it was actually observed after the cool outburst F1 of the symbiotic binary AG~Dra (Leedj{\"a}rv et al., 2016). 

\subsubsection{Neglecting of the \mbox{He\,{\sc i}} emission line flux}
In Leedj{\"a}rv et al. (2004, 2016) and Merc et al. (2017), we assumed that fluxes of the \mbox{He\,{\sc i}} $\lambda$\,4471\,\AA\, emission line are negligible compared to the \mbox{H$_{\beta}$} $\lambda$\,4861\,\AA\, fluxes and we omitted them in our determination of the WD's temperature in AG~Dra using the equation \ref{eq:origin}. Similar simplification was proposed by Sokoloski et al. (2006) for Z~And. Based on this premise, the equation \ref{eq:origin} was simplified to the form
\begin{equation}\label{eq:fluxes}
T_{\rm hot}\,({\rm in\,10^4\,K}) \approx 14.16 \sqrt{F_{4686} \over F_{\rm H_{\beta}}} + 5.13,
\end{equation}
where $F_{4686} = EW_{4686}\,F^{\rm cont}_{4686}$ and $F_{\rm H_{\beta}} = EW_{\rm H_{\beta}}\,F^{\rm cont}_{\rm H_{\beta}}$. We verified this assumption for AG~Dra on the basis of our analysis of 70 spectra and we can conclude that $F_{4471} \leq (0.02-0.05) F_{\rm H_{\beta}}$ (see table \ref{tablelong} in appendix). Since the spectral lines considered have similar wavelengths and their ratio is used, we simplified the equation \ref{eq:fluxes} using the assumption that $F^{\rm cont}_{4686}$ $\approx$ $F^{\rm cont}_{\rm H_{\beta}}$. It allowed us to use EWs instead of fluxes
\begin{equation}\label{eq:EWs}
T_{\rm hot}\,({\rm in\,10^4\,K}) \approx 14.16 \sqrt{EW_{4686} \over EW_{\rm H_{\beta}}} + 5.13.
\end{equation}
\begin{figure}[t]
	\vspace{-1.0 cm}
	\includegraphics[width=\linewidth]{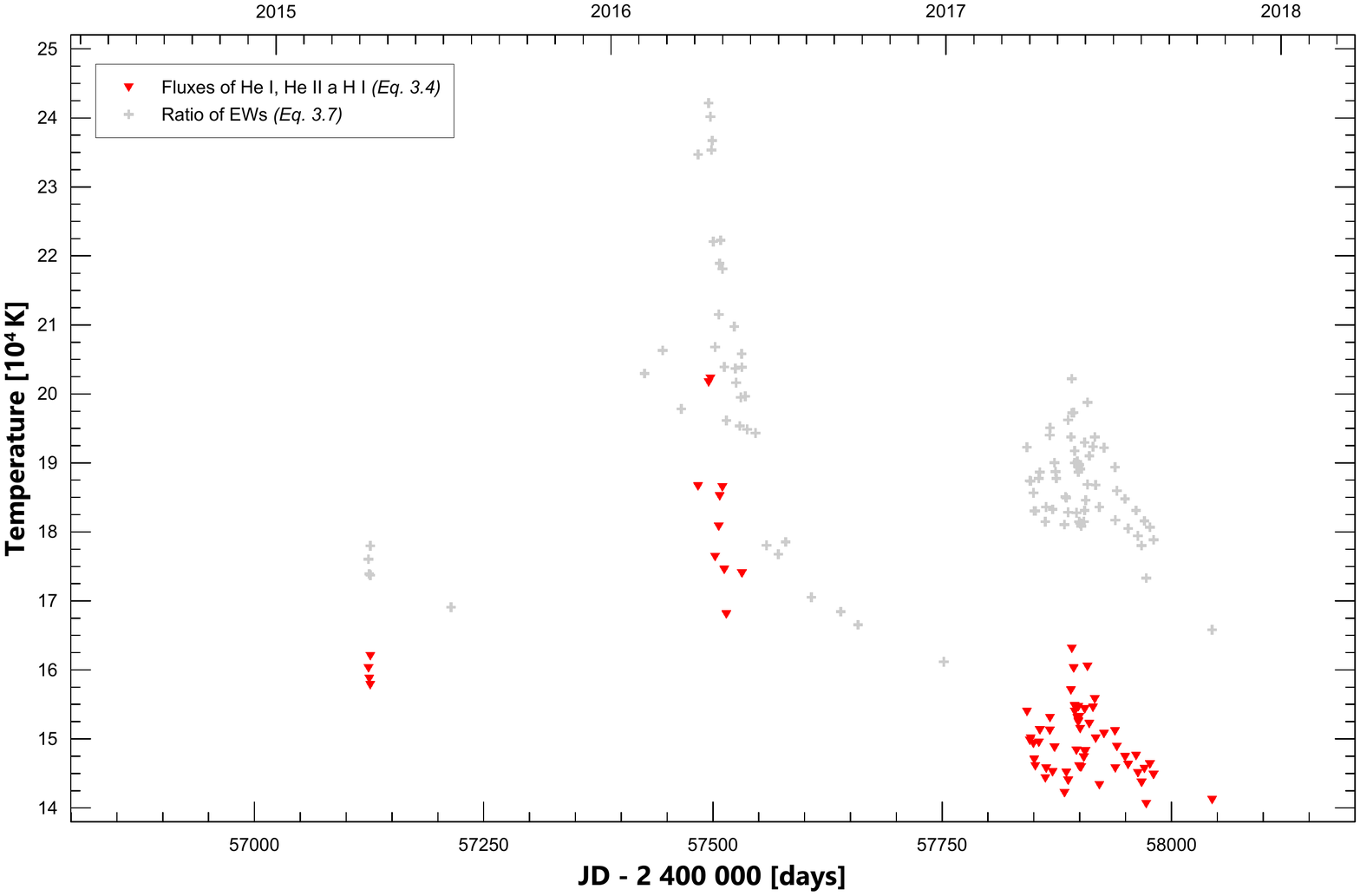}
	\caption{The effects of the \mbox{He\,{\sc i}} $ \lambda $\,4471\,\AA\, flux neglecting and of EWs use instead of fluxes for determination of the WD's temperature in AG~Dra. The temperatures computed using the equations \ref{eq:EWs} and \ref{eq:origin} are depicted by grey pluses and red triangles, respectively. For the latter values, the \mbox{He\,{\sc i}} and H$_{\beta}$ fluxes are also rectified for the orbital motion and contributions of the absorption component are removed before the calculation of the H$_{\beta}$ fluxes.}
	\label{fig:all_temp}
\end{figure}
On the other hand, if the ratio of $F^{\rm cont}_{4686}$ and $F^{\rm cont}_{\rm H_{\beta}}$ is determined by the average magnitudes of AG~Dra in the $B$ and $V$ filters with assumption of its extinction $E(B - V) = 0.0356 $, we obtained
\begin{equation}
\frac{F^{\rm cont}_{4686}}{F^{\rm cont}_{\rm H_{\beta}}} \approx 0.89.
\end{equation}
This result suggests that using of EWs instead of fluxes for determination of the WD's temperature increased its resulting values. Moreover, values of this ratio are changing markedly during the outbursts of AG~Dra. A combination of the two simplifications (neglecting of the \mbox{He\,{\sc i}} flux and using EWs instead of fluxes) increased in average the WD's temperature by around 7\%. It is worth note that for the particular spectra the increase varies from 5\% to 10\%. The temperatures of the hot component in AG~Dra derived using the equations \ref{eq:origin} and \ref{eq:EWs} are depicted in the figure \ref{fig:all_temp} and are listed in the table \ref{tablelong} in the appendix).

\section{Discussion and conclusions}

In the presented study, we investigated the WD's temperature and behaviour of the symbiotic binary AG~Dra during its ongoing active stage and compared them with those for previous stages of activity and quiescence. As we showed, to study the intrinsic variation of the WD's temperature (e.g. due to the outburst activity of AG~Dra), it is necessary to disentangle particular effects in observed changes of the studied emission lines EWs. We detailed the effects that can influence the resulting temperature estimates, and we were able to quantify the magnitude of each impact. The main findings can be summarised as follows.

\begin{itemize}
    \item The EWs of low excitation lines (e.g. \mbox{He\,{\sc i}} and \mbox{H\,{\sc i}}) depend on the orbital phase of AG~Dra. If the WD's temperature is derived using these measurements, its long-term course is apparently affected by the orbital motion of this symbiotic binary.
    \item It is necessary to subtract the orbital variations to investigate the WD temperature changes due to outburst activity of AG~Dra. For this purpose, we fitted the variation of the H$_{\beta}$ and \mbox{He\,{\sc i}} $\lambda$\,4471\,\AA\, EWs during quiescence stages by sinusoidal functions and the residuals after subtraction of the obtained fits from original values were used for analysis of the WD temperature behaviour.
    \item The absorption component of the H$_{\beta}$ emission line, which is always presented in spectra of AG~Dra and is the most prominent during and after the inferior conjunction of the giant in this symbiotic binary, reduces its EWs by 20-25\%.
    \item The presence of the H$_{\beta}$ absorbent component does not affect the course of the WD temperature variations but increases the temperature estimates by approximately 10\%.
    \item Neglecting of the \mbox{He\,{\sc i}} flux in the original Iijima (1981) method could result in the increase of the calculated WD's temperatures by around 7\% and this effect does not change their course significantly. 
\end{itemize}

Moreover, during the quiescence, the symbiotic nebula is only partly ionisation-bounded  (Leedj{\"a}rv et al., 2016 and Skopal et al., 2017 based on the results of Nussbaumer \& Vogel, 1987), so a part of ionising photons can escape from the nebula and the observed ratio ($F_{4686}$/$F_{\rm H_{\beta}}$)$_{\rm obs}$ > $F_{4686}$/$F_{\rm H_{\beta}}$. This effect leads to further overestimation of the hot component temperature. For completeness, we can add that the method based on the optical emission line properties can give an overrated estimate of the hot component temperature also in such case when other physical mechanisms than photoionisation of cool star wind are contributing to the emission strengths (e.g. collision shocks, disk coronae). The method moreover does not take into account the effect of the WD wind, that can also increase the equivalent widths of the emission lines of ionised helium and neutral hydrogen (Sion et al., 2012). In total, the WD temperature values should be overestimated by up to 15-20\%. 

We prepared the procedure to disentangle particular influences on the WD temperature estimations in order to study its intrinsic variation due to outburst activity of AG~Dra. The procedure can be summarised into three main steps: (i)\,We measured the EWs of the \mbox{He\,{\sc i}}, \mbox{He\,{\sc ii}} and H$_{\beta}$ emission lines. (ii) The EWs of the \mbox{He\,{\sc i}} and H$_{\beta}$ were rectified for the orbital motion and the influence of the absorption component was taken into account. (iii) The temperatures of the WD were calculated according to Iijima (1981) method (equation \ref{eq:origin}). The temperatures calculated using this procedure are listed in the table \ref{tablelong} in the appendix of this paper. 

\section*{Acknowledgements}
We are grateful to all ARAS members that contributed their observations to this paper, particularly we acknowledge and thank Christian Buil, Franck Boubault, Fran\c{c}ois Teyssier, Joan Guarro Flo, Tim Lester, Olivier Garde and Peter Somogyi. We acknowledge with thanks the variable star observations from the AAVSO International Database contributed by observers worldwide and used in this research. This work was supported by the Slovak Research and Development Agency grant No. APVV-15-0458 and the institutional research funding IUT 40-1 of the Estonian Ministry of Education and Research. Last but not least, JM would like to thank Franco Giovannelli for an invitation to the conference and for the support of his stay in Palermo.

\bigskip
\bigskip
\noindent {\bf DISCUSSION}

\bigskip
\noindent {\bf JOANNA MIKO\L{}AJEWSKA:} How one can trigger $\sim$360-day pulsations of a red giant with $\sim$35 solar radii?

\bigskip
\noindent {\bf JAROSLAV MERC:} In addition to the $\sim$550 days period related to the orbital motion another period $\sim$355 days have been detected both in photometry and spectroscopy of AG~Dra. The possible explanation as pulsations of the giant was proposed in the paper of Petr{\'i}k et al. (1998) and supported later by G{\'a}lis et al. (1999) and Friedjung et al. (2003), in which the radial velocities based on the giant's absorption lines measurements were analysed and this period and its relation to the giant was confirmed. Obviously, its explanation as pulsation period of the giant remains open (but there is no better explanation yet) and definitely, more theoretical work is required to better understand the exact nature of this period.

\bigskip
\noindent {\bf KENJI TANABE:} I have a graduate course student who is interested in symbiotic stars and also obtained some spectra of AG Dra. During the outburst, is the low resolution (R $\approx$ 400 - 600) sufficient, or higher resolution (R $\approx$ 2000) is favourable? What is the resolution of spectra you have used?

\bigskip
\noindent {\bf JAROSLAV MERC:} For the analysis of the behaviour of AG~Dra (EWs or the WD's temperature), we have used the spectra with medium and higher resolution (R $\approx$ 2000 - 11000). Low-resolution spectra have been used for determination of the outburst onset and should be used for modelling of the spectral energy distribution (e.g. see Skopal et al., 2017).

\appendix
\newpage
\section{Observational data}
\begin{center}
\small
{\setlength{\tabcolsep}{7pt}
\begin{longtable}[h!]{cccccccccc}
\captionsetup{width=\linewidth}
\caption{\normalsize The EW (in \AA\,) and flux (in 10$^{-12}$ erg\,s$^{-1}$\,cm$^{-2}$) measurements of the selected emission lines based on spectra of AG~Dra from the ARAS database. The WD's temperatures are listed in 10$^{4}$\,K. Earlier data from Tartu Observatory are available online (Leedj{\"a}rv et al., 2004, 2016).}
\label{tablelong}\\
\textbf{JD} & \textbf{Obs.*} & \textbf{EW$_{4471}$} & \textbf{F$_{4471}$} & \textbf{EW$_{4686}$} & \textbf{F$_{4686}$} & \textbf{EW$_{4861}$} & \textbf{F$_{4861}$} & \textbf{T$_{hot}$ $^1$} & \textbf{T$_{hot}$ $^2$    }\\
\hline
\endfirsthead

\multicolumn{10}{c}%
{{\normalsize \tablename\ \thetable{} -- continued from previous page}} \vspace{10pt}\\

\textbf{JD} & \textbf{Obs.*} & \textbf{EW$_{4471}$} & \textbf{F$_{4471}$} & \textbf{EW$_{4686}$} & \textbf{F$_{4686}$} & \textbf{EW$_{4861}$} & \textbf{F$_{4861}$} & \textbf{T$_{hot}$ $^1$} & \textbf{T$_{hot}$ $^2$}\\ \hline \endhead

\hline \multicolumn{10}{|r|}{{Continued on next page}} \\ \hline
\endfoot

\hline \hline
\endlastfoot

57124.52 & BUI & 2.78 & 0.44 & 18.80 & 6.53 & 24.22 & 9.36 & 17.18 & 16.02 \\
57125.35 & FMT & 2.74 & 0.41 & 17.80 & 6.18 & 23.74 & 9.18 & 17.00 & 15.87 \\
57126.38 & BUI & 3.14 & 0.44 & 18.71 & 6.49 & 25.03 & 9.68 & 16.89 & 15.78 \\
57126.52 & BUI & 3.08 & 0.45 & 18.45 & 6.40 & 23.05 & 8.91 & 17.43 & 16.20 \\
57214.39 & PSO & - & - & 16.43 & 5.65 & 23.76 & 9.13 & 15.54 & - \\
57425.53 & PSO & - & - & 18.25 & 6.00 & 15.91 & 5.88 & 16.71 & - \\
57445.39 & PSO & - & - & 16.34 & 5.48 & 13.63 & 5.12 & 17.03 & - \\
57465.47 & PSO & - & - & 16.03 & 5.45 & 14.97 & 5.71 & 16.79 & - \\
57483.73 & LES & 1.80 & 0.25 & 19.68 & 7.38 & 11.73 & 4.89 & 19.97 & 18.66 \\
57495.42 & FMT & 1.75 & 0.38 & 28.96 & 16.12 & 15.95 & 9.40 & 21.13 & 20.17 \\
57497.35 & FMT & 1.76 & 0.40 & 34.75 & 20.34 & 19.53 & 12.08 & 21.08 & 20.22 \\
57498.35 & PSO & - & - & 35.94 & 21.46 & 21.28 & 13.37 & 20.72 & - \\
57499.35 & PSO & - & - & 35.22 & 21.70 & 20.54 & 13.26 & 20.85 & - \\
57500.44 & PSO & - & - & 30.12 & 20.03 & 20.71 & 13.37 & 19.63 & - \\
57502.40 & FMT & 2.05 & 0.88 & 30.78 & 21.58 & 25.53 & 18.58 & 18.40 & 17.64 \\
57506.37 & FMT & 2.01 & 0.68 & 31.51 & 19.25 & 24.62 & 15.70 & 18.86 & 18.08 \\
57507.38 & FMT & 2.02 & 0.60 & 30.46 & 17.12 & 21.74 & 13.03 & 19.51 & 18.52 \\
57508.48 & PSO & - & - & 36.64 & 20.23 & 25.13 & 14.77 & 19.82 & - \\
57510.37 & FMT & 2.06 & 0.61 & 33.84 & 20.37 & 24.38 & 15.50 & 19.50 & 18.64 \\
57512.37 & FMT & 2.03 & 0.64 & 25.48 & 14.90 & 21.94 & 13.54 & 18.31 & 17.45 \\
57514.62 & LES & 2.03 & 0.80 & 20.83 & 15.97 & 19.90 & 16.23 & 17.69 & 16.80 \\
57523.36 & FMT & - & - & 34.84 & 17.27 & 27.81 & 14.84 & 18.97 & - \\
57524.31 & PSO & - & - & 38.93 & 19.29 & 33.62 & 17.94 & 18.41 & - \\
57525.38 & PSO & - & - & 37.68 & 17.81 & 33.44 & 17.09 & 18.25 & - \\
57529.39 & PSO & - & - & 34.30 & 16.36 & 33.14 & 17.09 & 17.75 & - \\
57530.33 & PSO & - & - & 34.61 & 16.51 & 31.60 & 16.29 & 18.14 & - \\
57531.38 & PSO & - & - & 36.42 & 17.10 & 30.60 & 15.78 & 18.71 & - \\
57531.70 & LES & 2.48 & 0.73 & 31.85 & 14.95 & 27.43 & 14.15 & 18.58 & 17.40 \\
57535.41 & PSO & - & - & 36.35 & 17.07 & 33.12 & 17.64 & 18.20 & - \\
57537.42 & PSO & - & - & 35.01 & 16.40 & 34.07 & 17.25 & 17.79 & - \\
57546.41 & PSO & - & - & 31.08 & 13.38 & 30.46 & 14.37 & 17.89 & - \\
57558.46 & PSO & - & - & 25.81 & 10.44 & 32.20 & 14.39 & 16.54 & - \\
57571.42 & PSO & - & - & 21.25 & 8.00 & 27.06 & 11.34 & 16.67 & - \\
57579.45 & FBO & - & - & 22.13 & 8.33 & 27.40 & 11.48 & 16.91 & - \\
57607.37 & PSO & - & - & 18.64 & 6.79 & 26.29 & 10.58 & 16.45 & - \\
57639.39 & PSO & - & - & 19.02 & 6.77 & 27.80 & 11.01 & 16.31 & - \\
57658.26 & PSO & - & - & 18.17 & 6.40 & 27.43 & 10.77 & 16.14 & - \\
57751.65 & PSO & - & - & 16.43 & 5.48 & 27.28 & 9.50 & 14.94 & - \\
57842.43 & OGA & 2.44 & 0.29 & 18.77 & 6.25 & 18.94 & 7.07 & 16.35 & 15.39 \\
57845.60 & JGF & 2.57 & 0.39 & 18.53 & 6.19 & 20.06 & 7.51 & 15.97 & 14.97 \\
57846.41 & FMT & 2.48 & 0.33 & 18.36 & 6.13 & 19.88 & 7.45 & 15.95 & 15.01 \\
57849.40 & OGA & 2.35 & 0.24 & 17.03 & 5.79 & 18.91 & 7.18 & 15.75 & 14.93 \\
57850.35 & FMT & 2.45 & 0.37 & 18.21 & 6.27 & 21.04 & 8.12 & 15.61 & 14.70 \\
57851.47 & JGF & 2.74 & 0.42 & 17.30 & 6.01 & 20.00 & 7.77 & 15.56 & 14.60 \\
57855.35 & FMT & 2.50 & 0.39 & 19.10 & 6.62 & 20.57 & 8.01 & 15.92 & 14.94 \\
57856.48 & OGA & 2.42 & 0.29 & 19.36 & 6.88 & 20.58 & 8.17 & 15.99 & 15.12 \\
57862.41 & FMT & 2.46 & 0.42 & 16.56 & 6.00 & 19.60 & 7.88 & 15.33 & 14.43 \\
57863.36 & FMT & 2.35 & 0.36 & 16.33 & 6.12 & 18.70 & 7.81 & 15.45 & 14.57 \\
57867.37 & FMT & 2.47 & 0.43 & 17.83 & 6.72 & 17.55 & 7.40 & 16.15 & 15.12 \\
57867.44 & JGF & 2.45 & 0.33 & 18.26 & 6.88 & 17.71 & 7.47 & 16.25 & 15.30 \\
57870.38 & FMT & 2.40 & 0.43 & 16.82 & 6.34 & 19.37 & 8.07 & 15.39 & 14.52 \\
57872.43 & JGF & 2.23 & 0.39 & 16.79 & 6.54 & 17.50 & 7.59 & 15.80 & 14.87 \\
57873.44 & PSO & - & - & 18.57 & 7.00 & 19.71 & 8.29 & 15.82 & - \\
57874.40 & PSO & - & - & 18.65 & 7.00 & 20.09 & 8.10 & 15.75 & - \\
57883.37 & FMT & 2.16 & 0.29 & 14.52 & 5.16 & 17.29 & 6.86 & 15.03 & 14.22 \\
57884.38 & PSO & - & - & 18.61 & 6.67 & 20.85 & 8.36 & 15.52 & - \\
57885.37 & JGF & 2.48 & 0.40 & 17.08 & 6.20 & 19.17 & 7.78 & 15.42 & 14.51 \\
57887.37 & FMT & 1.95 & 0.28 & 15.23 & 5.61 & 17.65 & 7.21 & 15.16 & 14.39 \\
57887.41 & PSO & - & - & 21.06 & 7.87 & 20.10 & 8.21 & 16.33 & - \\
57890.40 & JGF & 2.65 & 0.65 & 32.39 & 14.33 & 32.00 & 15.45 & 16.55 & 15.70 \\
57891.41 & JGF & 2.77 & 0.59 & 35.38 & 14.87 & 31.15 & 14.30 & 17.20 & 16.31 \\
57891.48 & PSO & - & - & 36.05 & 15.12 & 33.94 & 15.58 & 16.87 & - \\
57893.37 & JGF & 2.85 & 0.69 & 36.24 & 15.43 & 34.10 & 15.73 & 16.87 & 16.02 \\
57894.40 & JGF & 2.35 & 0.45 & 28.86 & 11.16 & 29.34 & 12.53 & 16.31 & 15.47 \\
57894.53 & OGA & 2.24 & 0.38 & 28.49 & 11.02 & 29.71 & 12.68 & 16.18 & 15.40 \\
57896.39 & JGF & 2.43 & 0.40 & 25.73 & 10.02 & 29.84 & 12.89 & 15.60 & 14.83 \\
57897.42 & JGF & 2.16 & 0.34 & 26.85 & 9.96 & 27.90 & 11.57 & 16.14 & 15.30 \\
57898.38 & JGF & 2.62 & 0.33 & 26.58 & 10.04 & 27.79 & 11.31 & 16.10 & 15.46 \\
57898.38 & JGF & 2.29 & 0.36 & 26.04 & 9.64 & 27.67 & 11.26 & 16.00 & 15.24 \\
57898.39 & FMT & 2.19 & 0.38 & 25.62 & 9.48 & 26.93 & 10.96 & 16.04 & 15.25 \\
57899.39 & FMT & 2.21 & 0.35 & 23.17 & 8.36 & 27.43 & 11.12 & 15.42 & 14.60 \\
57899.39 & JGF & 2.26 & 0.37 & 26.22 & 9.69 & 27.43 & 11.13 & 16.08 & 15.31 \\
57900.40 & JGF & 2.24 & 0.36 & 24.92 & 9.07 & 26.30 & 10.69 & 15.99 & 15.14 \\
57900.40 & FMT & 2.26 & 0.40 & 23.10 & 8.41 & 27.43 & 11.15 & 15.41 & 14.59 \\
57901.40 & PSO & - & - & 25.12 & 9.00 & 29.86 & 12.00 & 15.46 & - \\
57901.49 & JGF & 2.71 & 0.44 & 25.12 & 9.00 & 30.02 & 12.06 & 15.44 & 14.59 \\
57904.41 & FMT & 2.31 & 0.46 & 25.95 & 10.01 & 30.69 & 13.08 & 15.50 & 14.73 \\
57905.39 & FMT & 2.34 & 0.44 & 24.50 & 9.52 & 28.28 & 12.05 & 15.57 & 14.82 \\
57905.40 & JGF & 2.34 & 0.44 & 26.38 & 10.18 & 26.36 & 11.24 & 16.28 & 15.43 \\
57906.40 & FMT & 2.16 & 0.41 & 23.51 & 8.90 & 26.53 & 11.13 & 15.63 & 14.82 \\
57908.38 & PSO & - & - & 25.30 & 9.49 & 27.59 & 11.52 & 15.84 & - \\
57908.42 & JGF & 2.83 & 0.41 & 35.15 & 13.19 & 32.40 & 13.52 & 16.92 & 16.04 \\
57910.40 & JGF & 2.19 & 0.38 & 25.03 & 9.37 & 25.71 & 10.78 & 16.10 & 15.22 \\
57914.42 & JGF & 2.63 & 0.49 & 29.81 & 10.81 & 30.03 & 12.06 & 16.34 & 15.45 \\
57916.42 & JGF & 2.50 & 0.41 & 29.69 & 10.38 & 29.33 & 11.34 & 16.43 & 15.57 \\
57917.42 & FMT & 2.46 & 0.42 & 25.68 & 8.97 & 28.04 & 10.84 & 15.84 & 15.01 \\
57921.42 & FMT & 2.40 & 0.46 & 24.62 & 8.36 & 28.20 & 11.63 & 15.59 & 14.33 \\
57926.40 & JGF & 2.46 & 0.40 & 26.34 & 9.23 & 26.60 & 10.96 & 16.22 & 15.07 \\
57938.41 & JGF & 2.35 & 0.38 & 24.47 & 8.56 & 25.74 & 10.03 & 15.98 & 15.11 \\
57938.66 & LES & 2.63 & 0.37 & 22.53 & 7.88 & 26.56 & 10.42 & 15.40 & 14.57 \\
57940.40 & JGF & 2.42 & 0.37 & 24.07 & 8.29 & 26.61 & 10.26 & 15.74 & 14.88 \\
57949.42 & JGF & 2.56 & 0.45 & 24.53 & 7.86 & 27.61 & 9.99 & 15.70 & 14.74 \\
57952.71 & LES & 1.98 & 0.26 & 23.57 & 7.84 & 28.31 & 10.57 & 15.40 & 14.62 \\
57961.41 & JGF & 2.04 & 0.25 & 22.41 & 7.25 & 25.86 & 9.43 & 15.56 & 14.75 \\
57963.42 & JGF & 1.88 & 0.26 & 20.99 & 6.79 & 25.63 & 9.30 & 15.27 & 14.50 \\
57967.42 & JGF & 1.69 & 0.20 & 18.35 & 5.96 & 22.92 & 8.34 & 15.09 & 14.36 \\
57970.39 & JGF & 2.05 & 0.27 & 20.07 & 6.34 & 23.71 & 8.48 & 15.42 & 14.57 \\
57972.37 & FMT & 1.75 & 0.23 & 18.65 & 5.92 & 25.11 & 9.02 & 14.82 & 14.06 \\
57976.43 & JGF & 1.99 & 0.29 & 21.70 & 6.79 & 25.99 & 9.12 & 15.45 & 14.63 \\
57980.40 & JGF & 2.58 & 0.44 & 23.95 & 7.61 & 29.51 & 10.63 & 15.42 & 14.48 \\
58044.32 & FMT & 2.02 & 0.22 & 18.16 & 5.68 & 27.78 & 9.73 & 14.83 & 14.12               
\end{longtable}}
\end{center}
\vspace{-30pt}

\noindent \small \textbf{Notes:}\\
\noindent \small{* BUI - Christian Buil, FBO - Franck Boubault, FMT - Fran\c{c}ois Teyssier, JGF - Joan Guarro Flo, LES - Tim Lester, OGA - Olivier Garde, PSO - Peter Somogyi.}
\setlength{\leftskip}{0cm}

\noindent \small{$^1$ Computed using the equation \ref{eq:EWs}.}

\noindent \small{$^2$ Computed using the equation \ref{eq:origin}. The \mbox{He\,{\sc i}} and H$_{\beta}$ fluxes were rectified for orbital motion and contributions of the absorption component were taken into account before the calculation of the H$_{\beta}$ fluxes.}

\end{document}